\documentclass[prb,twocolumn,showpacs,preprintnumbers,amsmath,amssymb]{revtex4}

\usepackage{dcolumn}
\usepackage{bm}
\usepackage{epsfig}

\begin{document}


\title{Absence of local magnetic moments in Ru and Rh impurities and clusters on Ag(100) and Pt(997)}

\author{J. Honolka}
\email{j.honolka@fkf.mpg.de} \affiliation{Max-Planck-Institut f\"ur
Festk\"orperforschung, Heisenbergstrasse 1, 70569 Stuttgart,
Germany}
\author{K. Kuhnke}
\affiliation{Max-Planck-Institut f\"ur Festk\"orperforschung,
Heisenbergstrasse 1, 70569 Stuttgart, Germany}
\author{L. Vitali}
\affiliation{Max-Planck-Institut f\"ur Festk\"orperforschung,
Heisenbergstrasse 1, 70569 Stuttgart, Germany}
\author{A. Enders}
\affiliation{Max-Planck-Institut f\"ur Festk\"orperforschung,
Heisenbergstrasse 1, 70569 Stuttgart, Germany}
\author{S. Gardonio}
\affiliation{Istituto di Struttura della Materia, Consiglio
Nazionale delle Ricerche, Area Science Park, 34012 Trieste, Italy}
\author{C. Carbone}
\affiliation{Istituto di Struttura della Materia, Consiglio
Nazionale delle Ricerche, Area Science Park, 34012 Trieste, Italy}
\author{S. R. Krishnakumar}
\affiliation{International Centre for Theoretical Physics (ICTP),
Strada Costiera 11, 34100 Trieste, Italy}
\author{K. Kern}
\affiliation{Max-Planck-Institut f\"ur Festk\"orperforschung,
Heisenbergstrasse 1, 70569 Stuttgart, Germany}
\author{P.~Bencok}
\affiliation{European Synchrotron Radiation Facility, BP 200, 38043
Grenoble, France}
\author{S. Stepanow}
\affiliation{Centre d'Investigacions en Nanoci\`{e}ncia i
Nanotecnologia (CSIC-ICN), UAB Campus, 08193 Bellaterra, Spain}
\author{P. Gambardella}
\affiliation{Instituci\'{o} Catalana de Recerca i Estudis
Avan\c{c}ats (ICREA) \\ and Centre d'Investigacions en
Nanoci\`{e}ncia i Nanotecnologia (CSIC-ICN), UAB Campus, 08193
Bellaterra, Spain}

\date{\today}

\begin{abstract}
The magnetism of quench-condensed Ru and Rh impurities and metal
films on Ag(100) and Pt(997) has been studied using x-ray magnetic
circular dichroism. In the coverage range between 0.22~ML and
2.0~ML no dichroic signal was detected at the M$_{3,2}$ absorption
edges of Ru on Ag(100) at a temperature of 5~K in the presence of
an applied magnetic field. The same was found for coverages
between 0.12~ML and 0.5~ML of Rh on Ag(100) and Pt(997). It is
concluded that the magnetic moments of single impurities, small
clusters of various shape and monolayers of the 4$d$ metals are
below the detection limit of 0.04 $\mu_{\text B}$ per atom. These
results provide an unambiguous determination of the local magnetic
moment of Ru and Rh deposited on nonmagnetic transition-metal
surfaces, which are in contrast with theoretical predictions.
\end{abstract}


\maketitle

\section{Introduction}
The search for magnetism in 4$d$ metal elements has generated a
large number of theoretical and experimental investigations, often
in contradiction with each other. Stern-Gerlach experiments have
shown that Rh clusters of 10-30 atoms in molecular beams possess a
magnetic moment of about $1-0.5$~$\mu_{\text B}$ per atom, which
tends to vanish with increasing cluster size.\cite{cox93prl,ACox94}
This result has been generally explained by the reduced coordination
of the 4$d$ atoms, which narrows the width of the electronic bands
and increases the density of states at the Fermi level so as to
fulfill the Stoner criterium for the appearance of ferromagnetism.
Similar arguments applied to monolayer (ML) films and clusters
deposited on nonmagnetic substrates have stimulated \textit{ab
initio} theoretical efforts, which predicted 4$d$ magnetism for a
broad variety of nanostructures, including clusters of different
size and shape,\cite{Lang94,wildberger95prl,Step99,cabria02prb}
atomic chains,\cite{Baz00,bellini01prb} and mono- or
bilayers.\cite{eriksson91prl,zhu91prb,wu92prb,bluegel92prl,Tur95,Bluegel95}
According to systematic calculations by Eriksson \textit{et
al.}\cite{eriksson91prl} and Bl\"{u}gel,\cite{bluegel92prl}
overlayers of the late 4$d$ elements Rh and Ru  exhibit
ferromagnetism when placed on a Ag(100) substrate with magnetic
moments of $1.0$~$\mu_{\text B}$ and $1.7$~$\mu_{\text B}$ per atom,
respectively. Calculations of clusters with finite dimensions for
coverages below 1 ML predict that the 4$d$ magnetic moment
significantly depends on geometry and substrate: compact clusters as
well as elongated chain configurations on Ag(100) present finite
moments varying from 0.3 to 2.0~$\mu_{\text
B}$.\cite{wildberger95prl,Step99,cabria02prb,Baz00,bellini01prb} In
the case of isolated adatoms (impurities) both Ru and Rh are
expected to be magnetic on Ag(100), with local moments of
$2.2$~$\mu_{\text B}$ and $0.3$~$\mu_{\text B}$,
respectively.\cite{Lang94,wildberger95prl} In all these studies it
has been stressed that the values of the moments vary strongly with
the local coordination of 4$d$ atoms. Moreover, according to theory,
the intermixing of Rh and Ru layers with the Ag surface layers as
well as the growth of imperfect films with noninteger coverage
between 1 and 2~ML can strongly decrease 4$d$
magnetism,\cite{Tur95,Bluegel95} making its experimental
verification rather difficult.\newline \indent So far, most
experiments have concentrated on Rh systems. Magneto-optical Kerr
(MOKE) investigations of 1~ML Rh/Ag(100), \cite{mulhollan91prb}
0.5-5~ML Rh/Au(100),\cite{liu91prb} and 0 - 6~ML
Rh/Au(111)\cite{chado01prb} performed at temperatures down to 40,
100, and 30 K, respectively, indeed failed to confirm the presence
of ferromagnetism in Rh. Several explanations have been put forward
to reconciliate the MOKE results with theoretical calculations. It
was remarked that the impossibility to grow ideal 4$d$ monolayers,
i.e., the formation of a diffuse Rh-Ag interface as well as
three-dimensional island growth, could induce a significant
reduction of the Rh magnetism.\cite{Tur95,Bluegel95} Structural
imperfections, strain relaxation, and intermixing are typical
features of epitaxial films grown out of equilibrium that can
scarcely be taken into account by band structure models. Moreover,
as ab-initio calculations are usually carried out at 0~K, the lack
of ferromagnetism could be attributed to the limited temperature
range probed by MOKE experiments, specifically if the Curie
temperature is situated below 30-40 K. The existence of long-range
ferromagnetic order in 4$d$ metal layers, however, is not the only
prediction that can be experimentally tested. Prior to that, in
fact, one should prove that Rh and Ru atoms deposited on a noble
metal surface preserve part of their gas-phase local magnetic
moment. MOKE performed at relatively high temperature and low
magnetic fields ($\leq 0.2$ T) does not yield information in this
respect. Early photoemission measurements showed a splitting of the
Rh 4$s$ core-levels for 1-3 ML Rh/Ag(100), which was taken as an
indicator for the presence of a local magnetic moment.\cite{li91prb}
In a more recent study, Beckmann and Bergmann have investigated Rh
(Ru) impurities quench-condensed on Au (Au and Ag) films grown on
quartz supports by measuring the film magnetoresistance and
anomalous Hall effect.\cite{beckmann97prb} By modelling the
dephasing of electrons scattered from impurities, these authors
concluded that Ru impurities on Au and Ag films possess a small but
finite moment of about $0.4 \, \mu_B$ and suggested that Rh clusters
have a fluctuating moment the order of $0.1 \, \mu_B$. These
interesting results outlined a nontrivial magnetic behavior of 4$d$
metal overlayers, even though the estimate of the Ru and Rh
magnetization was performed in a nonlocal way, measuring the
electrical properties of the supporting
films.\cite{beckmann97prb,bergmann83prb}

In this work, we report on element-specific x-ray magnetic circular
dichroism (XMCD) measurements of the local magnetic moment of Ru and
Rh adatoms and cluster ensembles deposited at 5~K on Ag and Pt
surfaces. We show that no magnetic moment is detected in the
coverage range between 0.12 and 2.0 ML, independently of the
magnitude of externally applied static magnetic fields. These
results show unambiguously that Ru and Rh in the form of impurities,
small clusters, and quench-condensed films possess negligible local
magnetic moments under static field conditions, in contrast to
theoretical predictions. The XMCD data call either for a revision of
existing \textit{ab initio} models or for taking into account
many-body, temperature-dependent effects that can induce spin
fluctuations over timescales faster than the time resolution of the
present experiment.

\section{Experiment}

The measurements were carried out at beamline ID08 of the European
Synchrotron Radiation Facility (ESRF) in Grenoble.  Sample
preparation and magnetic characterization were performed {\it in
situ} under ultra high vacuum conditions. Single-crystal Ag(100) and
Pt(997) surfaces were cleaned by repeated sputter and annealing
cycles until a satisfactory low-energy electron diffraction pattern
was obtained and no contaminants were detected by x-ray absorption
spectroscopy (XAS). In order to prevent surface diffusion and
intermixing, the 4$d$ metals were deposited at a temperature of
$T=5$~K using an e-beam evaporator directly connected to the
XAS-XMCD vacuum chamber. Contamination checks prior and after the
measurements were done on the O XAS absorption line to exclude
possible contamination from oxygen, carbon monoxide, and water. Room
temperature STM was used, prior to the XMCD measurements, to further
verify the quality of the substrates and, after the XMCD
measurements, to estimate the total 4$d$ overlayer coverage.
Intermediate coverages were determined by linear extrapolation of
the evaporation time at constant evaporation rate, which was
monitored in real time by the current of ionized atoms reaching the
sample.
\begin{figure}
\epsfxsize=8.5cm
$$
\epsfbox{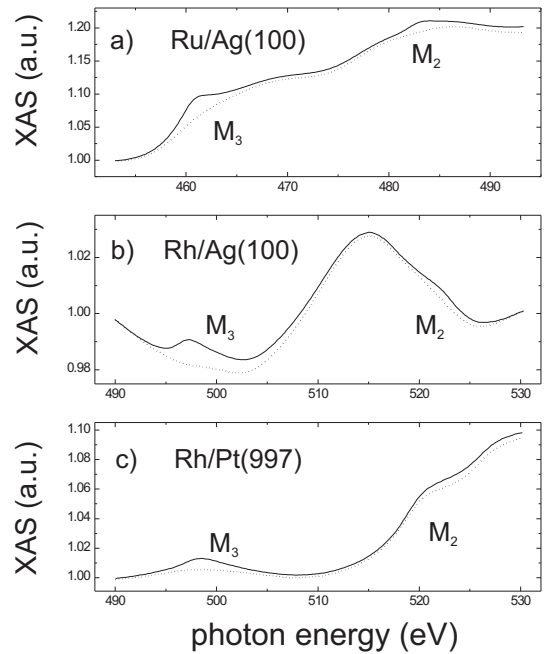}
$$
\caption{\label{XAS} $M_{3,2}$ XAS spectra of (a) Ru (0.22ML), (b)
Rh (0.09ML) deposited on Ag(100), and (c) Rh (0.12ML) on Pt(997) at
$T=5$~K. The spectra were normalized to unity at the M$_{3}$
pre-edge energies. The background XAS measured prior to deposition
on pristine Ag and Pt substrates is represented by dotted lines.}
\end{figure}
XAS spectra were taken in the region of the $M_{3,2}$ lines of Rh
(Ru), situated at energies of 497 eV (461 eV) and 522 eV (483 eV),
respectively. As shown in Figure~\ref{XAS} at low coverages the
$M_{3,2}$ lines are superimposed by a rather large background signal
from the substrates. The structures in the background intensity
originate from the Ag $M_{5,4}$ thresholds situated at lower energy,
while for Pt the increase of background intensity around 525~eV
stems from the $N_{3}$ absorption edge. The XAS intensity was
measured by recording the total photoelectron current by means of an
electrometer as a function of the x-ray energy, and positive
($\sigma^+$) or negative ($\sigma^-$) x-ray circular polarization
($99 \pm 1 \%$ polarization degree). The integration time at each
energy point was set to 0.3 s. Magnetic fields of up to $B=6$~T were
applied parallel and antiparallel to the photon beam. The angle of
incidence of the beam was varied between $\Theta = 0^{\circ}$
(normal incidence) and $\Theta = 55^{\circ}$ (oblique incidence) to
probe the out-of-plane and in-plane XMCD. In the following we refer
to the XAS signal as the average intensity
$(\sigma^{+}+\sigma^{-})/2$ and to the XMCD as
$(\sigma^{+}-\sigma^{-})$. The experimental timescales for measuring
spectra as well as ramping of the magnets to their designated values
are of the order of 10-100s.

\section{Results}
\label{results}

\begin{figure}
\epsfxsize=8.5cm
$$
\epsfbox{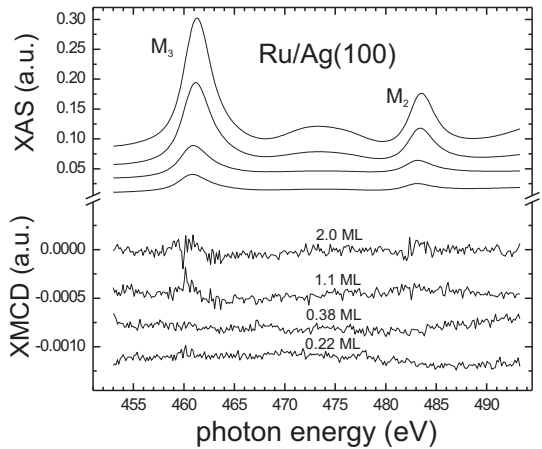}
$$
\caption{\label{XAS_Ru} XAS and XMCD spectra of Ru on Ag(100) for
coverages $(0.22\pm 0.05)$, $(0.38\pm 0.05)$, $(1.1\pm 0.1)$ and
$(2.0\pm 0.2)$ ML, measured at $T=5$~K and $B=5$~T. The XAS spectra
are background subtracted. For clarity the spectra have been offset
in the vertical direction.}
\end{figure}

\begin{figure}
\epsfxsize=8.5cm
$$
\epsfbox{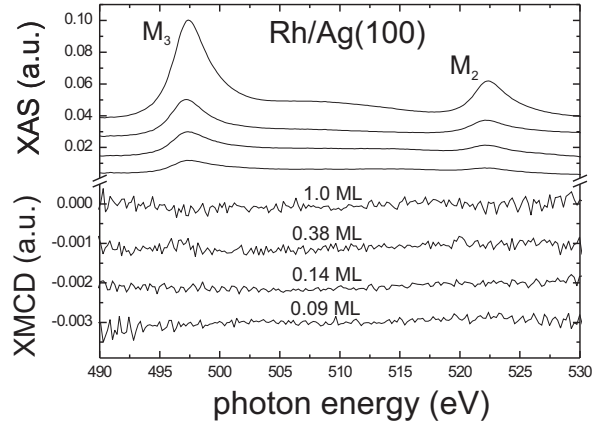}
$$
\caption{\label{XAS_RhAg} XAS and XMCD spectra of Rh on Ag(100) for
coverages of $(0.09\pm 0.05)$, $(0.14\pm 0.05)$, $(0.38\pm 0.1)$,
and $(1.0\pm 0.1)$ ML measured at $T=5$~K and $B=6$~T. The XAS
spectra are background substrated. For clarity the spectra have been
offset in the vertical direction.}
\end{figure}

\begin{figure}
\epsfxsize=8.5cm
$$
\epsfbox{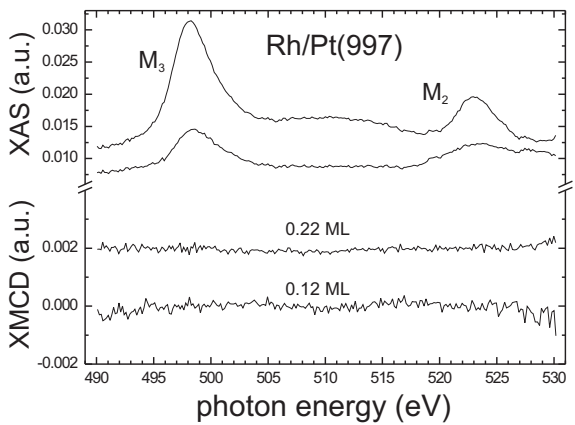}
$$
\caption{\label{XAS_RhPt} XAS and XMCD spectra of Rh on Pt(997) for
coverages of $(0.12\pm 0.05)$ and $(0.22\pm 0.05)$ ML measured at
$T=5$~K and $B=5$~T. The XAS spectra are background substrated. For
clarity the spectra have been offset in the vertical direction.}
\end{figure}

Figure~\ref{XAS_Ru} shows the XAS and XMCD spectra of 0.22 to 2.0 ML
Ru on Ag(100) recorded at $\Theta = 0^{\circ}$. The background XAS
of the clean Ag(100) substrate in the $M_{3,2}$ region
(Fig.~\ref{XAS}) was subtracted from the original XAS spectra in
order to show the residual Ru signal. The $M_{3}$ and $M_{2}$ edges
of Ru stand out clearly, together with a diffuse feature between the
two peaks, which is typical of both Ru and Rh.\cite{Tom98} In order
to gain magnetic information, the spectra were recorded in an
applied field of $B=5$~T at 5~K. In these conditions, even for
paramagnetic species, the presence of a magnetic moment on the 4$d$
metal atoms should produce nonzero dichroism with opposite peaks at
the $M_{3}$ and $M_{2}$ edges, as observed, e.g., in the case of
Fe-induced magnetism in Ru/Fe multilayers.\cite{Tom98,lin98prb} The
Ru XMCD spectra are instead flat within the noise limit, thus
showing that the Ru magnetic moment is below the sensitivity of our
experiment. XMCD spectra at $\Theta = 55^{\circ}$, not shown here,
give similar results compared to $\Theta = 0^{\circ}$ and will not
be commented further. Small features appearing in the XMCD around
the $M_{3,2}$ edges were proven to be of nonmagnetic origin, as they
do not change sign when the direction of the magnetic field is
reversed. These artifact features correspond to differences of the
order $\pm 0.3 \%$ of the XAS $M_{3}$ edge jump, and could be due to
small energy or beam drift during the experiment. According to Tomaz
\textit{et al}.,\cite{Tom98,lin98prb} 1~$\mu_{\text B}$ per Ru atom
produces an XMCD signal of about $8 \%$ with respect to the XAS
intensity. If we set our lower detection limit equal to the
intensity of the artifact features in the Ru XMCD spectra we
therefore get a sensitivity of about 0.04~$\mu_{\text B}$ per atom.
Independently, we can estimate the upper limit of the magnetic
moment assuming the validity of the XMCD sum rules as proposed by
Carra \textit{et al}.,\cite{Carra93}. From the XMCD noise level of
1$\times 10^{-4}$ a.u. and the integral over the $M_{3,2}$ XAS
signal at 0.22ML we get a value of 0.006~$\mu_{\text B}$ per Ru atom
and hole in the 4$d$ shell. Disregarding charge transfer effects we
get an upper limit of 0.02~$\mu_{\text B}$ per Ru atom. At higher
coverages the upper limit of the moment scales down with the
increasing XAS signal.\newline \indent The measurements in
Fig.~\ref{XAS_Ru} set a lower bound for the average Ru magnetic
moment at different coverages. However, as mentioned in the
introduction, details in the coordination of the 4$d$ elements are
believed to play an important role in the formation of magnetic
moments in 4$d$ metals. We therefore need to discuss the growth mode
of Ru and Rh at low temperature. Deposition of Rh on a clean Ag(100)
surface at room temperature is known to produce overlayers, which
are partially covered by or intermixed with Ag, but which are
pseudomorphic with the Ag(100) substrate.\cite{li91prb,chang96prb}
No surface reconstruction has been observed. Also the growth of Ru
on Ag(100) is expected to be pseudomorphic.\cite{wu92prb}
\begin{figure}
\epsfxsize=8.5cm
$$
\epsfbox{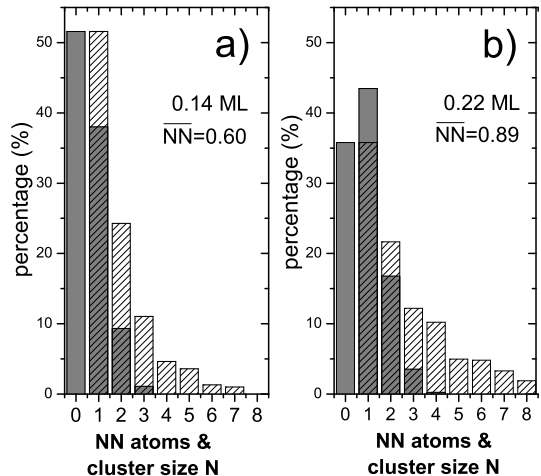}
$$
\caption{\label{statistics} Histogram of the number of nearest
neighbors (NN) per adatom on a (100) fcc two-dimensional lattice for
coverages of a) 0.14 and b) 0.22~ML assuming random adsorption (grey
bars). Hatched bars represent the percentage of atoms included in a
given cluster size N.}
\end{figure}
For the samples studied in this work, the low deposition temperature
of 5~K is expected to promote a random pseudomorphic growth with
inhibited surface diffusion and, at the same time, to suppress
thermally activated intermixing with Ag. Under these conditions, a
quantitative estimate of the degree of 4$d$-4$d$ coordination at a
given coverage can be done using a random occupation model where
each pseudomorphic (100) fcc adsorption site is occupied with a
probability equal to the coverage in ML units.
Figure~\ref{statistics} shows the statistics of the number of
nearest-neighbors (NN) per adatom for the measured Ru coverages of
0.14 and 0.22 ML (grey bars). In such a low-coverage regime
funneling events during deposition are expected to be rare, thus
providing a rather good estimation of the actual average
coordination of Ru atoms on the surface. In the figure, the NN
percentage is given by the grey bars regardless of the actual
cluster size N, while hatched bars represent the percentage of atoms
belonging to clusters of size N (The latter corresponds to N times
the percentage of clusters with size N). We see that the clusters
consist mainly of single impurities, dimers, and trimers, and the
majority of atoms is included in clusters with size $\leq 4$ atoms.
Note that a particular cluster size weights in the XAS intensity in
proportion to the percentage of the total number of atoms that it
contains. Thus, our detection limit of 0.02~$\mu_{\text B}$ per atom
has to be renormalized (increased) according to the hatched bars
percentages in Fig.~\ref{statistics} in the "worst-case" situation
where only one type of clusters presents a nonzero magnetization.
For single impurities in the 0.22 ML Ru sample this would mean an
upper limit of about 0.06~$\mu_{\text B}$, more than one order of
magnitude smaller than the predicted 2.2~$\mu_{\text B}$ per for
single Ru impurities,\cite{Lang94} 1.1-1.6~$\mu_{\text B}$ per Ru
atom in small clusters including dimers and trimers,\cite{Step99}
and 1.7-1.8~$\mu_{\text B}$ per Ru atom in an ideal monolayer on
Ag(100).\cite{Bluegel95, Tur95} We remark that one could in
principle deposit a lower amount of material ($~0.01$~ML) to address
uniquely single impurities. However, contrary to the 3$d$ elements
where XAS-XMCD at the $L_{3,2}$ edges is sensitive to coverages
$\leq 0.01$~ML,\cite{gambardella02prl,gambardella03science} the
lower cross-section for $M_{3,2}$ absorption together with the
strong substrate background in the energy region of interest limit
the XAS sensitivity to about 0.1~ML (Figs.~\ref{XAS_Ru},
\ref{XAS_RhAg}, \ref{XAS_RhPt}).

Measurements for Rh are shown in Fig.~\ref{XAS_RhAg} on the Ag(100)
surface and in Fig.~\ref{XAS_RhPt} on Pt(997) for coverages between
0.09 and 1.0~ML. All XMCD spectra have been measured at $T=5$~K and
$B=6$~T. No evidence for the presence of local Rh magnetic moments
is found. Similar considerations to those discussed for Ru based on
the noise level in the XMCD show that the detection limit at the
lowest coverage of $\sim$~0.1ML is about 0.02~$\mu_{\text B}$ per Rh
atom. Theory predicts magnetic moments of 0.3-1.0~$\mu_{\text B}$
per Rh atom in various cluster configurations with $N = 1,2,3,5,9$
and $21$ on
Ag(100),\cite{wildberger95prl,Step99,cabria02prb,Baz00,bellini01prb}
where $N$ is the number of Rh atoms in the cluster. These finite
moments are again clearly not observed in our experiment, which
should cover clusters sizes with $N = 1,2,3,4$, as shown in
Fig.~\ref{statistics}, and different configurations. The zero XMCD
of Rh on Pt(997), however, was expected on the basis of ab-initio
calculations of a nonmagnetic state for Rh impurities on
Pt(100).\cite{stepanyuk96prb} This is a general trend of 5$d$
\textit{vs.} 4$d$ substrates since, compared to Ag, the larger
extension of the Pt 5$d$ wavefunctions favors the hybridization
between adatoms and surface and reduces the impurity magnetic
moment.\cite{stepanyuk96prb}

\section{Discussion}
\label{discussion}

We discuss here the possible origin of the discrepancy between a
large number of consistent theoretical results and the missing local
moments in Ru and Rh observed by XMCD.\\
\emph{Alloying of 4$d$ metals with the substrate}. This is a widely
accepted argument for interpreting the absence of ferromagnetism in
Rh monolayers probed by MOKE.\cite{mulhollan91prb,liu91prb} STM,
Auger, thermal desorption, and ion scattering studies show that
ideal layer-by-layer growth of Rh and Ru on noble metal surfaces can
hardly be achieved at room temperature, due to the elemental
difference of surface free energy that promotes Rh-Ag
interdiffusion.\cite{chang96prb,schmitz89prb} Turek \textit{et al.}
calculated a substantial decrease of both the Ru and Rh magnetic
moment in subsurface positions for mixed Ru,Rh/Ag(100)
layers.\cite{Tur95} For small clusters, on the other hand, Stepanyuk
\textit{et al.} found an increase of the Ru and Rh moment upon
mixing with Ag.\cite{Step99}  In our samples, we expect alloying to
be suppressed due to deposition at 5~K. Even in the case of alloying
and reduced magnetization, however, the Ru and Rh moments should be
detectable as they are expected to be larger than our error
margins. \\
\emph{Relaxation effects in band-structure calculations}. Most
ab-initio calculations of the 4$d$ systems have been carried out
without fully relaxing the atomic lattice of the overlayer and the
substrate, often by assuming bulk-truncated pseudomorphic
configurations. When relaxation has been taken into account, this
was done by taking fixed average interlayer spacings.\cite{Tur95}
Wu and Freeman noted that relaxation shall not affect the tendency
of 4$d$ atoms to couple ferromagnetically.\cite{wu92prb} However,
due to the overlap of the electron wavefunctions, small
differences in the interatomic distance between the 4$d$ atoms and
the substrate, or between 4$d$ atoms in a
cluster,\cite{mokrousov06prl} could induce major changes in the
magnitude of the local magnetic moments. Such effects have not
been specifically addressed in theoretical treatments, perhaps for
the lack of related experimental information. Relaxing the Ru and
Rh position towards the substrate might partially or totally
quench the local 4$d$
moment. \\
\emph{Many-body effects}. Phenomena such as the Kondo effect or
local spin fluctuations might effectively reduce the magnetic moment
of impurities in metal hosts. While these effects have to be
considered in experiments carried out at finite temperature, they
are normally not included in \textit{ab-initio} density functional
treatments. For a fluctuating valence system with magnetic moment
averaging to zero, XMCD measurements are clearly too slow to provide
magnetic information. In such a case the motion of the magnetic
moment would no longer be dominated by the combined action of
thermal fluctuations and applied magnetic field, but by intrinsic
fluctuations due to interactions with the conduction electrons. The
magnetoresistance measurements by Beckmann and Bergmann seem to
support this hypothesis, showing that the dephasing rate of
electrons scattered off Ru and Rh impurities is suppressed at low
temperature as a result of screening of the local 4$d$
moments.\cite{beckmann97prb} As already noted by these authors,
however, Kondo screening and spin fluctuations are expected to
become less important for large clusters and therefore cannot
explain the missing magnetic moments in the whole coverage range of
the present investigation.

It is possible that more than one effect concurs to determine the
nonmagnetic state of Ru and Rh atoms in impurities and clusters
revealed by XMCD. With respect to structural relaxation,
experimental methods usually do not allow to determine the
interatomic distance between isolated impurities and the
substrate. However, detailed ab-initio calculations could address
this point. M{\"o}ssbauer spectroscopy on $^{99}$Ru isotopes could
provide information on the coordination state of Ru. STM
spectroscopy performed on individual impurities and clusters, on
the other hand, might reveal the role of Kondo screening in these
systems.\cite{WahlPRL}

\section{Conclusions}

We have performed element-specific measurements of the local
magnetic moment of the 4$d$ metals Ru and Rh deposited at 5~K on
Ag(100) and Pt(997). Ru was investigated in the coverage range from
0.22 to 2.0 ML on Ag(100). Rh was probed in the 0.09-1.0 ML coverage
range on Ag(100) and Pt(997) surfaces. No magnetic moments were
detected in both Ru and Rh impurities, clusters, and
quench-condensed layers within the detection limit of $0.02 \,
\mu_{\text B}$ per atom and in the presence of applied magnetic
fields of up to 6 T. These results show that the lack of
ferromagnetic order in Rh films observed by MOKE can be attributed
to quenching of the 4$d$ atomic moment upon deposition on
nonmagnetic noble metal substrates, even without taking into account
imperfect monolayer growth. The XMCD data are in contrast with
ab-initio density functional calculations: while Rh impurities on
Ag(100) represent a borderline magnetic-nonmagnetic system in
different theoretical
models,\cite{Lang94,wildberger95prl,Step99,cabria02prb} Rh clusters
containing a few atoms, Ru impurities, clusters, and monolayers are
predicted to have sizable magnetic moments of the order of
1~$\mu_{\text B}$ per
atom,\cite{Lang94,wildberger95prl,Step99,cabria02prb,bellini01prb,eriksson91prl,zhu91prb,wu92prb,bluegel92prl,Tur95,Bluegel95}
which are clearly not observed in the present experiment. The
magnetic behavior of 4$d$ elements deposited on nonmagnetic metal
surfaces appears to be a complex problem that cannot be entirely
treated in the framework of relaxation free, zero temperature
density functional models. On the experimental side, the possible
role of local spin fluctuations as a function of temperature and
cluster size remains to be clarified.

\section{Acknowledgments}

The authors acknowledge the European Synchrotron Radiation Facility
for provision of beamtime and thank Gilles Retout for technical
assistance in using beamline ID08.

\newpage 


\begin{thebibliography}{99}

\bibitem{cox93prl} A. J. Cox, J. G. Louderback,
and L. A. Bloomfield, Phys. Rev. Lett. 71, 923 (1993).

\bibitem{ACox94} A. J. Cox, J. G. Louderback, S. E. Apsel, and L. A.
Bloomfield, Phys. Rev. B 49, 12295 (1994).

\bibitem{Lang94}
P. Lang, V.S. Stepanyuk, K. Wildberger, R. Zeller, and P.H.
Dederichs, Sol. Stat. Comm. 92, 755 (1994)

\bibitem{wildberger95prl}
K. Wildberger, V. S. Stepanyuk, P. Lang, R. Zeller, and P. H.
Dederichs, Phys. Rev. Lett. 75, 509 (1995).

\bibitem{Step99}
V. S. Stepanyuk, W. Hergert, P. Rennert, K. Wildberger, R. Zeller,
and P. H. Dederichs, Phys. Rev. B 59, 1681 (1999).

\bibitem{cabria02prb}
I. Cabria, B. Nonas, R. Zeller, and P. H. Dederichs, Phys. Rev. B
65, 054414 (2002).

\bibitem{Baz00}
D.I. Bazhanov, W. Hergert, V. S. Stepanyuk, A. A. Katsnelson, P.
Rennert, K. Kokko, and C. Demangeat, Phys. Rev. B 62, 6415 (2000).

\bibitem{bellini01prb}
V. Bellini, N. Papanikolaou, R. Zeller, and P. H. Dederichs, Phys.
Rev. B 64, 094403 (2001).

\bibitem{eriksson91prl}
O. Eriksson, R. C. Albers, and A. M. Boring, Phys. Rev. Lett. 66,
1350 (1991).

\bibitem{zhu91prb}
M. J. Zhu, D. M. Bylander, and L. Kleinman, Phys. Rev. B 43, 4007
(1991).

\bibitem{wu92prb}
R. Wu and A. J. Freeman, Phys. Rev. B 45, 7222
(1992).

\bibitem{bluegel92prl} S. Bl\"{u}gel, Phys. Rev. Lett. 68, 851
(1992).

\bibitem{Bluegel95}
S. Bl\"{u}gel, Phys. Rev. B 51, 2025 (1995).

\bibitem{Tur95}
I. Turek, J. Kudrnovsk\'y, M. \v Sob, V. Drchal, and P. Weinberger,
Phys. Rev. Lett. 74, 2551 (1995).


\bibitem{mulhollan91prb}
G. A. Mulhollan, R. L. Fink, and J. L. Erskine, Phys. Rev. B 44
2393 (1991).

\bibitem{liu91prb}
C. Liu and S. D. Bader, Phys. Rev. B 44, 12062 (1991).

\bibitem{chado01prb}
I. Chado, F. Scheurer, and J. P. Bucher, Phys. Rev. B 64, 094410
(2001).

\bibitem{li91prb}
H. Li, S. C. Wu, D. Tian, Y. S. Li, J. Quinn, and F. Jona, Phys.
Rev. B 44, 1438 (1991).

\bibitem{Weaver74}
H. T. Weaver, and Rod K. Quinn, Phys. Rev. B 10, 1816 (1974).

\bibitem{beckmann97prb}
H. Beckmann and G. Bergmann, Phys. Rev. B 55, 14350 (1997).

\bibitem{bergmann83prb}
G. Bergmann, Phys. Rev. B 28, 2914 (1983)

\bibitem{Tom98} M.A. Tomaz, T. Lin, G. R. Harp, E. Hallin, T. K. Sham, and W. L. O'Brien, J. Vac. Sci. Technol. A, 16,
1359 (1998)

\bibitem{lin98prb}
T. Lin, M. A. Tomaz, M. M. Schwickert, and G. R. Harp, Phys. Rev.
B 58, 862 (1998).

\bibitem{Carra93}
P. Carra, B.T. Thole, Massimo Altarelli, Xingdong Wang, Phys. Rev.
Lett. 70, 694 (1993)


\bibitem{chang96prb}
S.-L. Chang, J.-M. Wen, P. A. Thiel, S. G\"unther, J. A. Meyer, and
R. J. Behm, Phys. Rev. B 53, 13747 (1996).

\bibitem{gambardella02prl}
P. Gambardella, S. S. Dhesi, S. Gardonio, C. Grazioli, P.
Ohresser, and C. Carbone, Phys. Rev. Lett. 88, 047202
(2002).

\bibitem{gambardella03science} P. Gambardella, S. Rusponi,
M. Veronese, S. S. Dhesi, C. Grazioli, A. Dallmeyer, I. Cabria, R.
Zeller, P. H. Dederichs, K. Kern, C. Carbone, H. Brune, Science
300, 1130 (2003).


\bibitem{tomaz97prb} A calibration of the Rh $M_{3,2}$ XMCD signal vs. magnetic moment is given in, e.g., M.A. Tomaz, D. C. Ingram, G. R. Harp, D. Lederman, E. Mayo, and W.
L. O'Brien, Phys. Rev. B 56, 5474 (1997).

\bibitem{stepanyuk96prb}
V. S. Stepanyuk, W. Hergert, K. Wildberger, R. Zeller, and P. H.
Dederichs, Phys. Rev. B 53, 2121 (1996).


\bibitem{schmitz89prb}
P. J. Schmitz, W.-Y. Leung, G. W. Graham, and P. A. Thiel, Phys.
Rev. B 40, 11477 (1989).

\bibitem{mokrousov06prl}
Y. Mokrousov, G. Bihlmayer, S. Heinze, and S. Bl\"ugel, Phys. Rev.
Lett. 96, 147201 (2006).

\bibitem{WahlPRL}
P. Wahl, L. Diekh\"oner, M. A. Schneider, L. Vitali, G. Wittich, and
K. Kern, Phys. Rev. Lett. 93, 176603 (2004)

\end{thebibliography}
\end{document}